\documentclass[12pt,a4paper]{article}
\usepackage[]{epsfig}         
\usepackage[T1]{fontenc}      
\usepackage[latin1]{inputenc} 
\usepackage[english]{babel}    
\usepackage[]{latexsym}       

\def\zfo{${{\mbox{$\bigcirc$}}\!\!\!\!\!\!\:{\mbox{2}}\,}$} 
\def\ffo{${{\mbox{$\bigcirc$}}\!\!\!\!\!\!\:{\mbox{5}}\,}$} 
      
\begin{document}
\thispagestyle{empty}
\title{Quasicrystals: Atomic coverings and windows are dual projects.}
\author{Peter Kramer\\
Institut f\"ur Theoretische Physik der Universit\"at\\
T\"ubingen, Germany.}

\maketitle

\section*{Abstract.}
In the window approach to quasicrystals, the atomic position space 
$E_{\parallel}$ is embedded
into a space $E^n = E_{\parallel} + E_{\perp}$. Windows  are attached 
to points of a lattice $\Lambda \in E^n$ . For standard 5fold and 
icosahedral tiling models, the windows  are perpendicular 
projections of dual Voronoi and Delone cells from $\Lambda$.
Their cuts by the position space $E_{\parallel}$ mark tiles and 
atomic positions. 
In the alternative covering approach, the position space
is covered by overlapping copies of a quasi-unit cell which 
carries a fixed atomic
configuration. The covering and window approach
to quasicrystals are shown to be dual projects:
$D$- and $V$- clusters are defined as 
projections to position space $E_{\parallel}$ 
of Delone or Voronoi cells. 
Decagonal $V$-clusters in the Penrose tiling, related to the decagon covering,
and two types of pentagonal
$D$-clusters in the triangle tiling of 5fold point symmetry 
with their windows are analyzed.
They are linked, cover 
position space and have definite windows. For functions compatible with 
the tilings they form domains of definition.
For icosahedral tilings the $V$-clusters are 
Kepler triacontahedra, the $D$-clusters are two icosahedra and one dodecahedron.

\section{Introduction:  Windows versus coverings.}

The standard approach for quasicrystal structure uses the project of 
{\em atomic
windows}:  These are polytopes with centers located at points of the 
unit cell from
a lattice $\Lambda$ in an embedding space $E^n = E_{\parallel}+ E_{\perp}$ . 
Atoms are then located on position space $E_{\parallel}$
by parallel cuts through
the windows, compare Katz and Gratias \cite{KA1, KA2} for icosahedral examples. 
The alternative project of {\em coverings} in quasicrystals was
introduced by Gummelt \cite{GU}, based on earlier concepts due to 
Burkov \cite{BU}
and Conway \cite{CO}. The Penrose tiling was interpreted in \cite{GU} as a system of 
decagonal covering clusters with overlaps.  These clusters were 
related to the concept of a {\em quasi-unit cell} by Jeong, Steinhardt et al.  
\cite{ST1, ST2, ST3}.  
The full quasicrystal structure is then composed from overlapping atomic
configurations on copies of the quasi-unit cell. Moreover, Steinhardt et al. interpret
these atomic configurations from the point of view of local 
energy. This interpretation takes up earlier work by Janot \cite{JA}. For 
a comment we refer to Urban \cite{UR}.
\vspace{0.2cm}  

What is the relation of the projects of coverings and windows to one another?  
What are the covering  clusters and links for a given tiling? What is the meaning of a (quasi-)
unit cell of a quasicrystal whose points, in contrast to crystals, 
are not related by the action of a
translation group?
In the present paper we try to answer these structure questions.  
We confront both
projects from the point of view of an embedding space, a lattice, its 
Voronoi cells $V$, its dual Delone cells $D^a, D^b, \ldots$ \cite{CO2}, 
and corresponding tilings.  The Voronoi and Delone cells of the lattice 
in the embedding space are the basis for a set of  
canonical tilings \cite{KR1}: The windows for these tilings  
are perpendicular projections of Voronoi or Delone cells.

We shall show that windows
and coverings are dual projects.
We define and analyze {\em  clusters as projections
to position space of Delone or Voronoi cells} $D^a_{\parallel}, D^b_{\parallel}, \ldots$
or $V_{\parallel}$. We term them {\em $D$-clusters or $V$-clusters}
respectively. We  determine {\em domains of definition for functions
on quasiperiodic tilings} and show that they can be related to 
sets of $V$-clusters 
or $D$-clusters.  These concepts are
introduced and illuminated in section 2 with the Fibonacci tiling, projected from the square
lattice, and in section 3 with a tiling projected from the lattice $A_2$.  In
section 4 we determine for the Penrose-Robinson tiling, projected 
from the lattice $A_4$ \cite{BA}, its V-clusters and obtain
the decagon covering of Gummelt \cite{GU}. In section 5 we construct
for the triangle tiling, dually projected from the lattice $A_4$ \cite {BA},
two pentagonal
$D^a, D^b$-clusters and the covering.
In section 6 we sketch the clusters for dual
icosahedral tilings. The $V$-clusters from the primitive and $F$-lattice are
Kepler triacontahedra, the three $D$-clusters from the 
$F$-lattice are icosahedra or dodecahedra.  
 
\section{Clusters  from  Delone  cells  in  the  Fibonacci tiling.}

\subsection*{Fibonacci tiling and klotz construction.}

Consider the Fibonacci tiling constructed from the square lattice   
$\Lambda$ of edge
length $s$ in $E^2$ by duality \cite{KR2}.  Its Voronoi cells $V(q)$ are 
squares centered at all
lattice points $q$.  Its dual Delone cells $D$ are squares centered at all
vertices  of the Voronoi cells.  All Delone cells belong to a single
translation orbit. A 2D fundamental domain ${\cal F}$ in $E^2$  
under the action of $\Lambda$ is provided by a single Voronoi cell $V$ or,
equivalently, by a single Delone cell $D$. The 1-boundaries $P$ of a Voronoi cell
are its four edge lines. The dual 1-boundaries $P^{*}$ of a Delone cell are its four
edge lines.  Pairs $P$, $P^{*}$ of dual 1-boundaries intersect in midpoints of the
edges of the squares.  Define the decomposition 
$E^2 = E_{\parallel} + E_{\perp}$  in the usual
fashion: $x_{\parallel}$ runs, w.r.t. to a densest lattice plane 

of $\Lambda$ along a line of slope
$\tau^{-1},\, \tau = (1 +\sqrt{5})/2$ respectively . 
The {\em klotz construction} \cite{KR4} for the Fibonacci tiling \cite{KR2} arises
as follows:
For each intersecting pair $P$, $P^{*}$, form at its intersection point the
convex klotz cells  $P_{\perp} \oplus P^{*}_{\parallel}$. The two klotz cells
$(A, B)$, see Fig.1,
are two squares $(A, B)$ of edge length $|L| =\tau |S|$,
$|S| =s/\sqrt{
\tau+2}$
respectively, with boundaries perpendicular or parallel to $E_{\parallel}$.
Any line with
points $x = x_{\parallel} + c_{\perp}$, $-\infty < x_{\parallel} < \infty$ intersects the klotz construction in a
Fibonacci tiling ${\cal T}^*$  with the tiles $(L := A_{\parallel}, 
S := B_{\parallel})$. A window for the full
local isomorphism class of all tilings ${\cal T}^*$  may be taken as a
perpendicular interval of length $|L| + |S|$ centered at a lattice point $q$.
This interval is the perpendicular
projection $V_{\perp}(q)$ of the Voronoi cell 
and appears in the klotz construction at all positions of lattice points $q$.

\subsection*{Quasiperiodic functions and fundamental domains.}

Atomic densities or electronic potentials  in quasicrystals 
require functional analysis 
on the position space. We recall the following relations on $E^n$: 
Under the geometric group action of 
$q \in \Lambda, x, x' \in E^n, (q, x) \rightarrow x'= x+q$,
a {\em fundamental domain} is a subset ${\cal F} \in E^n$ which has 
exactly one point from each translation orbit. The geometric group 
action yields for functions $f$ on $E^n$ the 
group operators $T_q: f(x) \rightarrow (T_qf)(x) := f(x-q)$.
Suppose now that $f$ is periodic on $E^n$ modulo $\Lambda$. 
We define a fundamental domain for $f$ as a subset ${\cal F} \in E^n$
such that any value of $f$ on $E^n- {\cal F}$ is obtained 
by the group action. Clearly a  fundamental 
domain for a periodic function $f$ can be identified  
with  a {\em fundamental domain} for the geometric action of $\Lambda$.

{\bf 1 Prop.} Two klotz cells  $(A,B)$ form a {\em fundamental domain 
${\cal F}$ for
functions $f$ on $E^2$ periodic modulo} $\Lambda$.

{\em Proof}: The pairs of dual boundaries underlying the cells $(A, B)$ 
are representatives of different translation orbits under $\Lambda$.
The cells do not overlap and together have the same volume as the Voronoi
square.  

\begin{center}
\begin{picture}(0,0)%
\epsfig{file=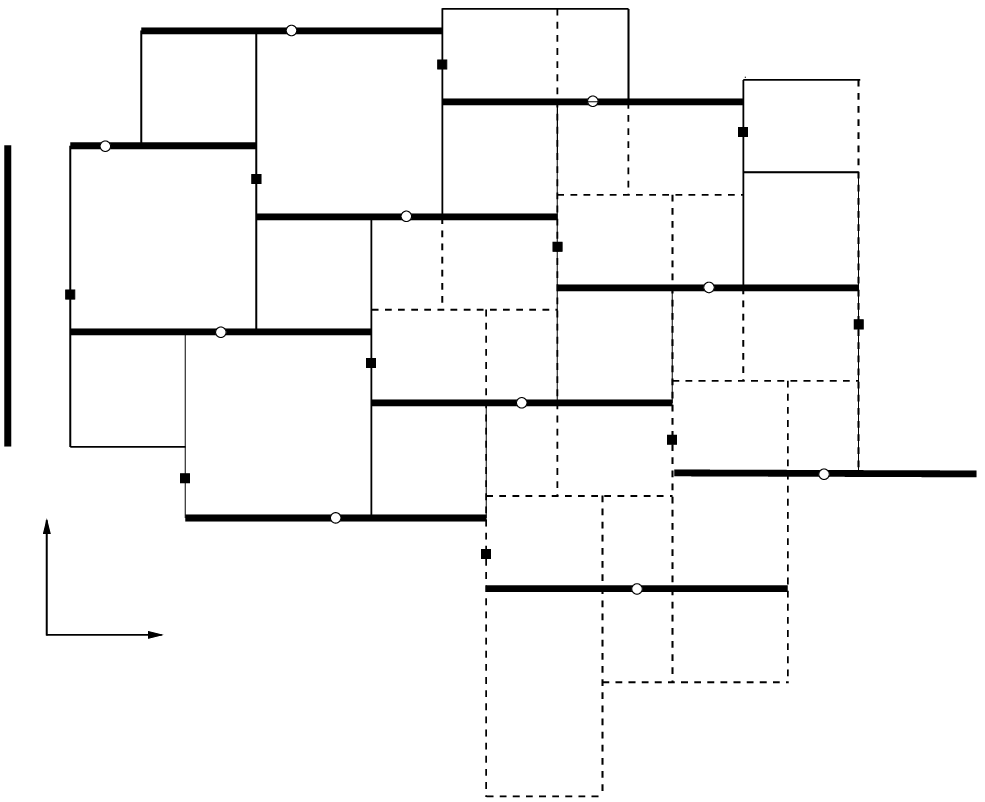}%
\end{picture}%
\setlength{\unitlength}{2072sp}%
\begingroup\makeatletter\ifx\SetFigFont\undefined%
\gdef\SetFigFont#1#2#3#4#5{%
  \reset@font\fontsize{#1}{#2pt}%
  \fontfamily{#3}\fontseries{#4}\fontshape{#5}%
  \selectfont}%
\fi\endgroup%
\begin{picture}(8964,7379)(4272,-16914)
\put(4456,-11581){\makebox(0,0)[lb]{\smash{\SetFigFont{12}{14.4}{\familydefault}{\mddefault}{\updefault}$V_{\perp}$}}}
\put(7156,-9691){\makebox(0,0)[lb]{\smash{\SetFigFont{12}{14.4}{\familydefault}{\mddefault}{\updefault}$D_{\parallel}$}}}
\put(5626,-11806){\makebox(0,0)[lb]{\smash{\SetFigFont{12}{14.4}{\familydefault}{\mddefault}{\updefault}$A$}}}
\put(5311,-13201){\makebox(0,0)[lb]{\smash{\SetFigFont{12}{14.4}{\familydefault}{\mddefault}{\updefault}$B$}}}
\put(5941,-10456){\makebox(0,0)[lb]{\smash{\SetFigFont{12}{14.4}{\familydefault}{\mddefault}{\updefault}$B$}}}
\put(7246,-10726){\makebox(0,0)[lb]{\smash{\SetFigFont{12}{14.4}{\familydefault}{\mddefault}{\updefault}$A$}}}
\put(9046,-15406){\makebox(0,0)[lb]{\smash{\SetFigFont{12}{14.4}{\familydefault}{\mddefault}{\updefault}$A'$}}}
\put(9901,-14686){\makebox(0,0)[lb]{\smash{\SetFigFont{12}{14.4}{\familydefault}{\mddefault}{\updefault}$B'$}}}
\put(4770,-14545){\makebox(0,0)[lb]{\smash{\SetFigFont{12}{14.4}{\familydefault}{\mddefault}{\updefault}$x_{\perp}$}}}
\put(5662,-15276){\makebox(0,0)[lb]{\smash{\SetFigFont{12}{14.4}{\familydefault}{\mddefault}{\updefault}$x_{\parallel}$}}}
\end{picture}

\end{center}

Fig. 1 The square lattice $\Lambda$ of edge length $s$ in $E^2$ has Voronoi 
squares $V(q)$, 
centered at lattice points $q$ (full squares), and Delone squares $D$, 
centered at 
Voronoi vertices (open circles). 
The lattice admits a periodic tiling into
two squares $(A, B)$ of edge lengths $|L| = \tau |S|, |S|$, called klotz cells 
and shown on the 
left-hand side.
The edges of these squares run along directions $x_{\parallel}$ horizontal, 
$x_{\perp}$ vertical of
slope $\tau^{-1}, \tau$  with respect to a densest lattice line.  
A pair $(A, B)$ of
two such squares form a fundamental domain ${\cal F}$  for the lattice. The
intersection of a parallel line with the two squares $(A, B)$ 
forms a Fibonacci tiling
${\cal T}^*$  with tiles $L = A_{\parallel}, S = B_{\parallel}$. The window of
the tiling is  $V_{\perp}(q)$ centered at lattice points $q$. Its size 
is indicated by a perpendicular bar on the left-hand side.
The Delone projections  $D_{\parallel}$ to position space $E_{\parallel}$ 
centered at Voronoi vertices (heavy lines at open circles)
provide  fundamental domains for functions compatible with the Fibonacci tiling. 
They 
bound pairs $A \cup B$ and $B \cup A$ from below and above. On parallel
line sections they give rise to $D$-clusters $(LS)$ or $(SL)$.
A second periodic tiling in $E^2$ 
with two
rectangles $(A', B')$ is shown on the lower right-hand side (dashed lines).  Its intersection with a
horizontal line $x = x_{\parallel} + c_{\perp}$, 
$-\infty < x_{\parallel} < \infty$, 
yields a deflated Fibonacci tiling ${\cal T}^*_{\tau^{-1}}$
with tiles $(L', S')$ of length scaled by the factor $\tau^{-1}$. 
The union of the
two tilings is shown in the middle part. In the parallel subtiling 
from this union, any cluster $(LS), (SL)$ of ${\cal T}^*$  gets 
the symmetric subdivision  $(L'S'L')$, and 
consecutive clusters are disjoint or linked by a tile $L'$.
\vspace{1cm}

{\em Quasiperiodic functions on a parallel line section} are characterized as follows:
Take a function $f$, defined by its values
on the two cells $(A, B)$ (or on any other equivalent fundamental domain) 
${\cal F}$,
and repeated on $E^2$ modulo $\Lambda$.  The restriction of $f$ to its values on a line
$x = x_{\parallel} + c_{\perp}, -\infty < x_{\parallel} < \infty$ 
gives rise to a quasiperiodic function on this
line.   The {\em fundamental domain 
for a quasiperiodic function} that determines its
value everywhere on the 1D horizontal line 
is seen in the embedding space $E^2$ as a
2D fundamental domain w.r.t. the action of $\Lambda$. 

Quasiperiodic functions of this general type do not take the same values
on different passages of the line through $A$ or through $B$, and so they
are not compatible with the Fibonacci tiling ${\cal T}^*$.
The class of quasiperiodic functions $f$ {\em compatible with the tiling} 
${\cal T}^*$ on the line $E_{\parallel}$ must have
the following restricted property, as discussed for example in  \cite{KR2}: 
On each of the two chosen klotz cells $(A, B)$,
its values must be {\em independent of} $x_{\perp}$.  These values by repetition 
under $\Lambda$ 
generate on any parallel line section a quasiperiodic function which takes the
same values on each passage through $A$ or $B$. We refer to Arnold \cite{AR}
for a discussion of quasiperiodic functions along similar lines.

{\bf 2 Def.}
Let the  tiling ${\cal T}$ consist of a minimal finite set 
$\langle p_i \rangle \in E_{\parallel} $ of prototiles $p_i$ 
and their translates appearing in ${\cal T}$.
A {\em fundamental domain for a quasiperiodic function $f$ on  
$E_{\parallel}$ compatible
with the tiling} is a subset of points $x_{\parallel} \in E_{\parallel}$  
which contains one and only one translate in ${\cal T}$ of any point from any 
prototile.
A fundamental domain with this property will be denoted by  
${\cal F}{({\cal T},\Lambda)}$. 
It depends  on  the
tiling ${\cal T}$, the lattice $\Lambda $, and the projection 
$E^n = E_{\parallel} + E_{\perp}$. The volume of the fundamental
domain is $|{\cal F}{({\cal T},\Lambda)}| = \sum_i |p_i|$. 

For the Fibonacci
tiling we find:

{\bf 3 Prop.}
The fundamental domain ${\cal F}{({\cal T}^*,\Lambda)}$ for any function 
$f$ compatible with the Fibonacci tiling ${\cal T}^*$ , 
can be taken as a line
interval in $E_{\parallel}$ of length $|L| + |S|$, 
consisting of a short and a long interval of the tiling ${\cal T}^*$.
The values of $f$ on the two intervals are then extended on each klotz cell 
$(A, B)$ to a 2D
function independent of $x_{\perp}$. By repetition modulo $\Lambda$  
and intersection with a parallel line they give rise to a 
particular quasiperiodic function.

\subsection*{Linked $D$-Clusters.}

The parallel projections $D_{\parallel}$ of the Delone squares are 
line sections of length $|L|+|S|$. These projections appear in the klotz 
tiling at the Delone centers.  Each one separates a
pair $(B, A)$ on top from a pair $(A, B)$ of klotz cells at the bottom. 
The boundary line itself we assign for uniqueness to the top pair of klotz
cells. If a
horizontal intersection line passes the top or bottom pair, any one of 
the two cuts provides a fundamental domain ${\cal F}{({\cal T}^*, \Lambda)}$.  
Both the (SL) and the (LS)
combination we term a $D$-{\em cluster}.

{\bf 4 Prop.} The Fibonacci tiling $({\cal T}^* , (L, S))$ is equivalent to a 
chain of linked
$D$-clusters of type $(LS), (SL)$. The clusters can be locally 
determined: Each
one is equivalent to the parallel projection $D_{\parallel}$ of a Delone cell and forms a fundamental
domain ${\cal F}{({\cal T}^*,\Lambda)}$. Consecutive clusters are disjoint or 
linked by a tile $S$ in the
form $(L(S)L)$.

{\em Proof}:  Compare Fig.1.  In the tiling ${\cal T}^*$ , form disjoint clusters from all
consecutive strings (LS) except for  the string $(LSLLS)$.
This string is interpreted with three clusters as $(L(S)L)(LS)$, with the
first two clusters linked by the tile $S$.

\subsection*{Symmetric subtiling and windows for  $D$-clusters.}  

The $D$-clusters as parallel projections $D_{\parallel}$ carry the
two alternative subtilings $(LS), (SL)$. We can remove this asymmetry 
by use of the deflated subtiling ${\cal T}^*_{\tau^{-1}}$. Its new
rectangular klotz cells $(A', B')$ in $E^2$ are shown with dashed lines 
in the lower part of Fig.1. 
The deflated tiles from parallel sections are $(L', S') = \tau^{-1}(L, S)$.
In the union of the original and deflated tiling, all projections 
$D_{\parallel}$ and hence all $D$-clusters get the symmetric subtiling $(LS), (SL) \rightarrow
(L'S'L')$. From the point of view of the deflated tiling, all strings 
$(L'L')$ separate disjoint clusters, all strings 
$(S'L'S')$ mark consecutive clusters linked by a tile $L'$.
Note that the $D$-clusters are not fundamental domains with repect
to the deflated tiling! So far we have not implemented the action of the 
point group,  generated here by inversion, on functions $f$. This 
would require restricting their domain to the counterpart of the asymmetric
unit in the terminology of periodic crystallography. 

The deflated tiling ${\cal T}^*_{\tau^{-1}}$ allows us to determine the
{\em window for centers of $D$-clusters}: The centers correspond to the 
midpoints of its tiles $S'$. Therefore their windows are the 
projections $B'_{\perp}$ of length $|L|$ of the rectangles 
$B'$, centered at the vertices of Voronoi cells in Fig. 1.

\section{$V$-Clusters in  a  tiling  with the lattice $A_2$.}

The klotz construction uses the duality between Voronoi and Delone cells for
a single lattice.  For illustration of duality we choose the root lattice 
$A_2$ in
$E^2$.  Its Voronoi cells $V$  are hexagons centered at lattice points, 
its Delone cells are two types of triangles $D^a,D^b$ centered at 
two different translation classes of vertex points of $V$. These cells
are shown in the top part of Fig. 2. 

\begin{center}
\begin{picture}(0,0)%
\epsfig{file=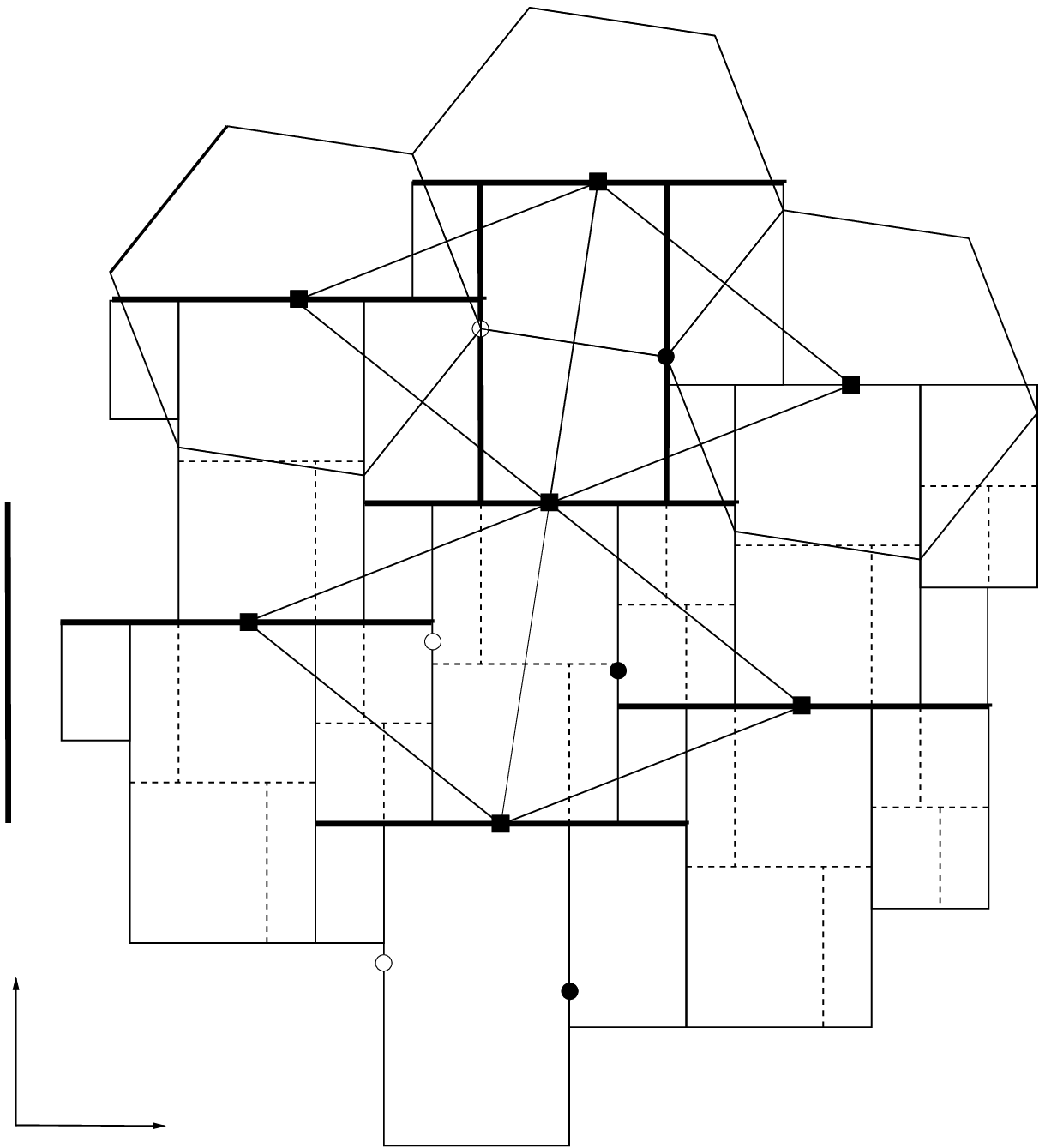}%
\end{picture}%
\setlength{\unitlength}{2072sp}%
\begingroup\makeatletter\ifx\SetFigFont\undefined%
\gdef\SetFigFont#1#2#3#4#5{%
  \reset@font\fontsize{#1}{#2pt}%
  \fontfamily{#3}\fontseries{#4}\fontshape{#5}%
  \selectfont}%
\fi\endgroup%
\begin{picture}(11055,12181)(3459,-13794)
\put(8820,-3185){\makebox(0,0)[lb]{\smash{\SetFigFont{12}{14.4}{\familydefault}{\mddefault}{\updefault}$V_{\parallel}$}}}
\put(8640,-5735){\makebox(0,0)[lb]{\smash{\SetFigFont{12}{14.4}{\familydefault}{\mddefault}{\updefault}$D^a_{\perp}$}}}
\put(10664,-6624){\makebox(0,0)[lb]{\smash{\SetFigFont{12}{14.4}{\familydefault}{\mddefault}{\updefault}$C$}}}
\put(9457,-6617){\makebox(0,0)[lb]{\smash{\SetFigFont{12}{14.4}{\familydefault}{\mddefault}{\updefault}$B$}}}
\put(7747,-6617){\makebox(0,0)[lb]{\smash{\SetFigFont{12}{14.4}{\familydefault}{\mddefault}{\updefault}$A$}}}
\put(6974,-11125){\makebox(0,0)[lb]{\smash{\SetFigFont{12}{14.4}{\familydefault}{\mddefault}{\updefault}$C$}}}
\put(8339,-11125){\makebox(0,0)[lb]{\smash{\SetFigFont{12}{14.4}{\familydefault}{\mddefault}{\updefault}$B$}}}
\put(9989,-11125){\makebox(0,0)[lb]{\smash{\SetFigFont{12}{14.4}{\familydefault}{\mddefault}{\updefault}$A$}}}
\put(3825,-12211){\makebox(0,0)[lb]{\smash{\SetFigFont{12}{14.4}{\familydefault}{\mddefault}{\updefault}$x_{\perp}$}}}
\put(4935,-13321){\makebox(0,0)[lb]{\smash{\SetFigFont{12}{14.4}{\familydefault}{\mddefault}{\updefault}$x_{\parallel}$}}}
\put(9721,-4471){\makebox(0,0)[lb]{\smash{\SetFigFont{12}{14.4}{\familydefault}{\mddefault}{\updefault}$D^b_{\perp}$}}}
\end{picture}

\end{center}

Fig. 2. The root lattice $A_2$. Its Voronoi cells $V(q)$ are hexagons
centered at lattice points $q$ (full squares), its 
Delone cells $D^a,D^b$ form  two  translation orbits of triangles centered 
at vertices of $V$ (open and full circles). 
They are shown in the top part.
The three rectangular klotz cells $A, B, C$ shown in the middle and 
bottom part arise from a 
decomposition $E^2 = E_{\parallel}
+ E_{\perp}$. The tiling ${\cal T}$ has the tiles $A_{\parallel},
B_{\parallel}, C_{\parallel}$. Its window consists of two projections 
$D^a_{\perp}, D^b_{\perp}$ centered at vertices of $V$. The size of the window
is indicated on the left-hand side by a perpendicular bar. 
The parallel projections $V_{\parallel}$ of Voronoi cells (heavy lines) 
at lattice points $q$ bound triples $A \cup B \cup C$ from below and 
$C \cup B \cup A$ from above.  On parallel
line sections they produce the $V$-clusters as
$(A_{\parallel}B_{\parallel}C_{\parallel})$ or
$(C_{\parallel}B_{\parallel}A_{\parallel})$. Both form domains of definition
for quasiperiodic functions $f$ compatible with the tiling. Consecutive 
$V$-clusters are linked by tiles $A_{\parallel}$ or $C_{\parallel}$.
A subdivision of the klotz cells $A, B$ is marked in the middle part by 
dashed lines. It yields an inversion-symmetric identical subtiling of the
two clusters.
\vspace{1cm}

Two dual tilings arise from a fixed decomposition 
$E^2 = E_{\parallel}+E_{\perp}$: In ${\cal T}$ the windows are projections
$D^a_{\perp},D^b_{\perp}$ of Delone cells, in ${\cal T}^*$ they are
projections $V_{\perp}$ of Voronoi cells. We choose ${\cal T}$ since
it is in analogy to the projection of the Penrose tiling in section 4. 
Three klotz
cells $A,B,C$ are shown at the bottom of Fig. 2. Their projections form the
tiles
$A_{\parallel},B_{\parallel}, C_{\parallel}$. The parallel projections 
$V_{\parallel}$ of Voronoi cells centered at lattice points yield one type
of  $V$-clusters as the strings
$(A_{\parallel}B_{\parallel}C_{\parallel})$ or
$(C_{\parallel}B_{\parallel}A_{\parallel})$. Both form {\em domains of 
definition} ${\cal F}{({\cal T}, A_2)}$ for quasiperiodic functions $f$ 
compatible with the tiling. Consecutive  $V$-clusters are linked by the 
tiles $A_{\parallel}$ or $C_{\parallel}$. 
The klotz cells $(A, B)$ may again be
subdivided by dashed lines as indicated in the middle part of Fig. 2.
This subdivision provides a subtiling of  both $V$-clusters
which is  symmetric under local inversion.  

\section{Decagonal $V$-clusters in the Penrose-Robinson tiling from the lattice $A_4$.}

We first summarize the construction of the decagon covering  due to 
Gummelt \cite{GU}. It uses the Penrose tiling ${\cal T}$ with rhombus  
edge length 
$s$ and Robinson decomposition together with the deflated tiling
${\cal T}_{\tau^{-1}}$
with edge length $\tau^{-1} s$ again with Robinson decomposition.
Select in ${\cal T}_{\tau^{-1}}$ all the vertex configurations 
{\bf king} and mark in each one  a center point, Fig. 3. 
From the empire of {\bf king}
it can be shown that a decagon of edge length $s$ is forced around the
center point, with a unique subdivision called the cartwheel \cite{CO}.
It is shown by Gummelt \cite{GU} that 
these cartwheel  decagons yield a {\em covering} which is equivalent to the Penrose
tiling. 

We employ the deflation sequence of tilings 
${\cal T}_{\tau} \rightarrow  {\cal T} \rightarrow {\cal T}_{\tau^{-1}}$,
compare also \cite{ST1},
to  redescribe the decagons according to Fig.3. We start with a thick rhombus in the first
tiling. On it we mark 
a   point with a full square. 
The first deflation yields in ${\cal T}$ the vertex configuration {\bf jack}. 
The next deflation yields in  ${\cal T}_{\tau^{-1}}$ the vertex 
configuration {\bf king}. This {\bf king} in Robinson subdivision enforces 
the cartwheel decagon of edge 
length $s$. The marked point is maintained in the three steps. 
It follows
from this sequence of deflations that {\em the decagon centers are fixed 
at the marked points on all the thick rhombus tiles of}
${\cal T}_{\tau}$.

Turn to the method of windows and projection.
We follow \cite{BA} and project the Penrose tiling 
of type ${\cal T}$ from the
root lattice $A_4$. The embedding space $E^4$ for this lattice splits into two 2D spaces
$E_{\parallel}, E_{\perp}$ of 5fold symmetry.  
There are four Delone cells whose perpendicular projections up to inversion
form a small and a large  pentagon. Their centers form four 
translation orbits of Voronoi vertices. The Voronoi vertices are called
the deep and shallow holes \cite{CO2} in the lattice $\Lambda$.
We denote representatives of the two shapes of Delone cells by $D^a,D^b$
and their projections by $D^a_{\perp},D^b_{\perp}$. Their
centers are shallow and deep holes respectively. Together with their 
mirror images they form the 
windows for the Penrose tiling. The klotz cells are formed from
$10$ pairs of dual 2D boundaries which represent different translation 
orbits. Similar as in the Fibonacci projection it can be shown that these 
$10$ klotz cells form a fundamental domain under $A_4$.
The Penrose tiles  are rhombic projections 
$P$ of 2D boundaries from the Voronoi cell. Their duals $P^*$
are acute and obtuse triangles,  projections
of 2D boundaries from the Delone cells. A {\em vertex configuration}
of Penrose tiles is coded by an overlap of the dual coding triangles inside
a Delone window \cite{BA}.
In the notation of \cite{BA}, the vertex configurations $1-3$ are coded in the 
small pentagon $D^a_{\perp}$, the vertex configurations $5-8$ 
in the large pentagon $D^b_{\perp}$. 
Turn then to the projections $V_{\parallel}$ and to the $V$-clusters.

{\bf 5 Prop.} The (linked) $V$-clusters of the Penrose-Robinson 
tiling are 
decagonal projections $V_{\parallel}$ to position space of edge length $s$. 
Two decagons form a fundamental domain
${\cal F}{({\cal T}, A_4)}$.

The projection is in shape equal to the window $V_{\perp}$ of the
dual triangle tiling, see section 5. For the Penrose tiling, 
a fundamental domain ${\cal F}({\cal T}, A_4)$ 
according to Def. 2 should contain 10 thick and 10 thin rhombus tiles. 
A single decagonal  $V_{\parallel}$-cluster of edge length $s$ can cover $5$ thick and 
$5$ thin rhombus tiles. Therefore not one, but two decagonal 
$V_{\parallel}$-clusters, rotated by an angle $2\pi/10$ to one another, 
are required to form  ${\cal F}({\cal T}, A_4)$. Both orientations occur
in the decagon covering and from Fig. 4 are related to the orientations 
of thick rhombus tiles from $({\cal T}_{\tau}, A_4)$.

We wish to locate the centers of these decagons by the projection method. 
This can be done with the
help of selected vertex configurations as follows: In the coding of the 
vertex configurations inside the small and large pentagons 
$D^a_{\perp}, D^b_{\perp}$, 
we look for {\em vertex configurations from coding triangles which 
share a single 
vertex  of} $D^a_{\perp}$ or $D^b_{\perp}$. Dualization of boundaries
\cite{KR4} 
implies that the tiles of such a vertex configuration 
{\em must belong to a single Voronoi cell}. This  condition holds true for 
the vertex configuration
2, the  {\bf jack} coded in $D^a_{\perp}$, and for vertex configuration 6,
the {\bf king} coded in $D^b_{\perp}$. Any king in the tiling forces a 
jack with the same
projected lattice point $q_{\parallel}$, and so we can restrict the analysis to jacks.

\begin{center}
\begin{picture}(0,0)%
\epsfig{file=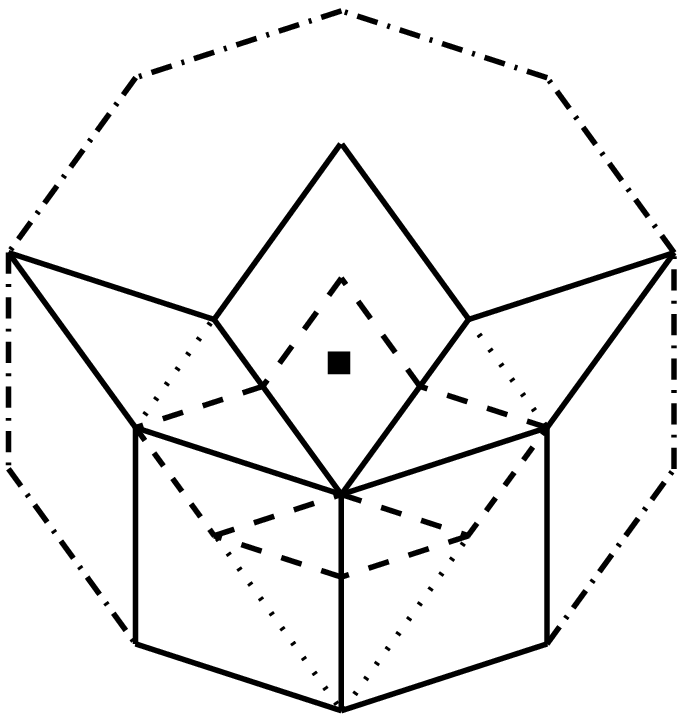}%
\end{picture}%
\setlength{\unitlength}{3315sp}%
\begingroup\makeatletter\ifx\SetFigFont\undefined%
\gdef\SetFigFont#1#2#3#4#5{%
  \reset@font\fontsize{#1}{#2pt}%
  \fontfamily{#3}\fontseries{#4}\fontshape{#5}%
  \selectfont}%
\fi\endgroup%
\begin{picture}(3870,4070)(771,-3443)
\end{picture}

\end{center}

Fig. 3. The Penrose tiling $({\cal T}, A_4)$, its inflation by $\tau$: 
$({\cal T}_{\tau}, A_4)$ and its deflation 
by $\tau^{-1}$: $({\cal T}_{\tau^{-1}}, A_4)$. 
At  a projected point $q_{\parallel}$ (full square) from the lattice $A_4$,
a thick rhombus tile of $({\cal T}_{\tau}, A_4)$ (dotted lines)
converts first into  a jack of $({\cal T}, A_4)$ (full lines) and  
then into a king of $({\cal T}_{\tau^{-1}}, A_4)$ (dashed lines). 
The empire of the king in Robinson decomposition forces the cartwheel 
decagon (-.- line) of Conway and
Gummelt. The decagons are $V$-clusters, two of them form a fundamental domain 
for functions $f$ compatible with the Penrose tiling.
\vspace{1cm}

We have then arrived from the 
$V$-clusters in the projection method at the decagons of Gummelt 
\cite{GU} with the following qualifications:

{\bf 6 Prop.}

(i) The $V$-clusters $V_{\parallel}$ of the Penrose tiling of edge length
$s$, located at projected lattice points $q_{\parallel}$ in the
vertex configurations {\bf jack}, coincide with the decagons of
Gummelt \cite{GU}.

(ii) Two decagons of edge length $s$ form a fundamental domain ${\cal F}{({\cal T}, A_4)}$ 
for functions $f$ compatible
with the Penrose tiling of the same edge length , not with any of its 
inflations or deflations.

(iii) The decagon edges around projected lattice points 
$q_{\parallel}$ on {\bf jacks},
Fig. 4, do not always appear 
as part of the rhombus tiling. 
The different decagonal domains of the covering become identical 
with the cartwheel upon subdivision by deflation to  
$({\cal T}_{\tau^{-1}}, A_4)$.

(iv) The decagon centers are fixed to projected lattice points $q_{\parallel}$ 
on the thick rhombus tiles of
the inflated tiling $({\cal T}_{\tau}, A_4)$. Their windows are,
up to perpendicular shifts from the rhombus vertex to the lattice point,
the acute triangles inside the Delone windows for this tiling.

\begin{center}
\begin{picture}(0,0)%
\epsfig{file=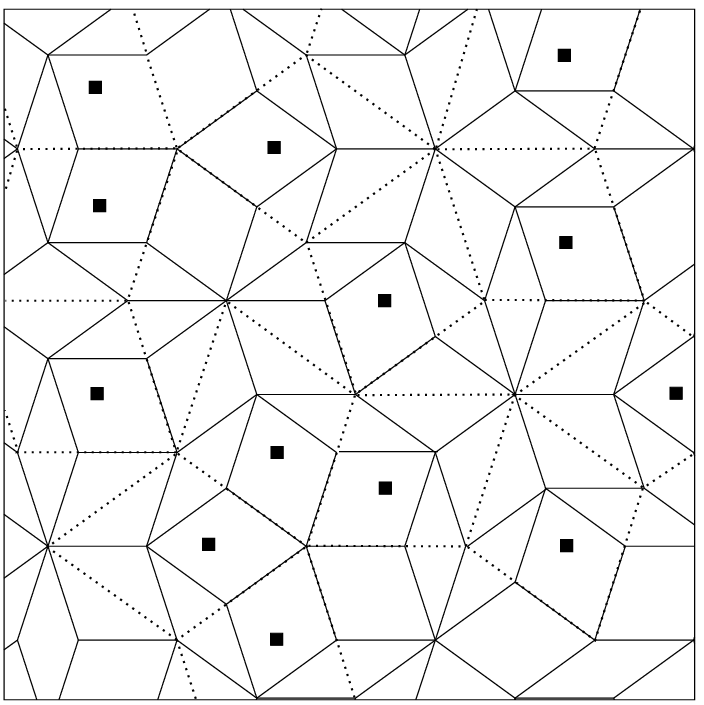}%
\end{picture}%
\setlength{\unitlength}{2368sp}%
\begingroup\makeatletter\ifx\SetFigFont\undefined%
\gdef\SetFigFont#1#2#3#4#5{%
  \reset@font\fontsize{#1}{#2pt}%
  \fontfamily{#3}\fontseries{#4}\fontshape{#5}%
  \selectfont}%
\fi\endgroup%
\begin{picture}(5570,5579)(3208,-6035)
\end{picture}

\end{center}
 
Fig. 4. Part of a Penrose tiling. Projected lattice points 
$q_{\parallel}$ are  marked by full squares. They are 
centers of decagonal $V$-clusters and cartwheel decagons at jack vertex
configurations, 
compare Fig. 3. These points are also located on the thick
rhombus tiles (dotted lines) of the inflated Penrose tiling.
\vspace{1cm}
 
The links between $V$-clusters are clearly related to  the sharing
of (parallel projected) dual boundaries. The window technique  
allows to characterize the linkage and the
relative frequencies and to compare with \cite{GU}.

\section{Pentagonal $D$-clusters in the triangle tiling from the lattice 
$A_4$.}

The triangle tiling \cite{BA} is the dual tiling ${\cal T}^*$ from the lattice 
$A_4$. The Voronoi and Delone cells are the same as for the Penrose 
tiling, but the projections interchange their role. 
Its window is a decagon $V_{\perp}$. Its tiles are an acute and an obtuse 
golden  triangle, coded in $E_{\perp}$ by Penrose rhombus tiles. 
There are $9$ vertex configurations. 

\begin{center}
\begin{picture}(0,0)%
\epsfig{file=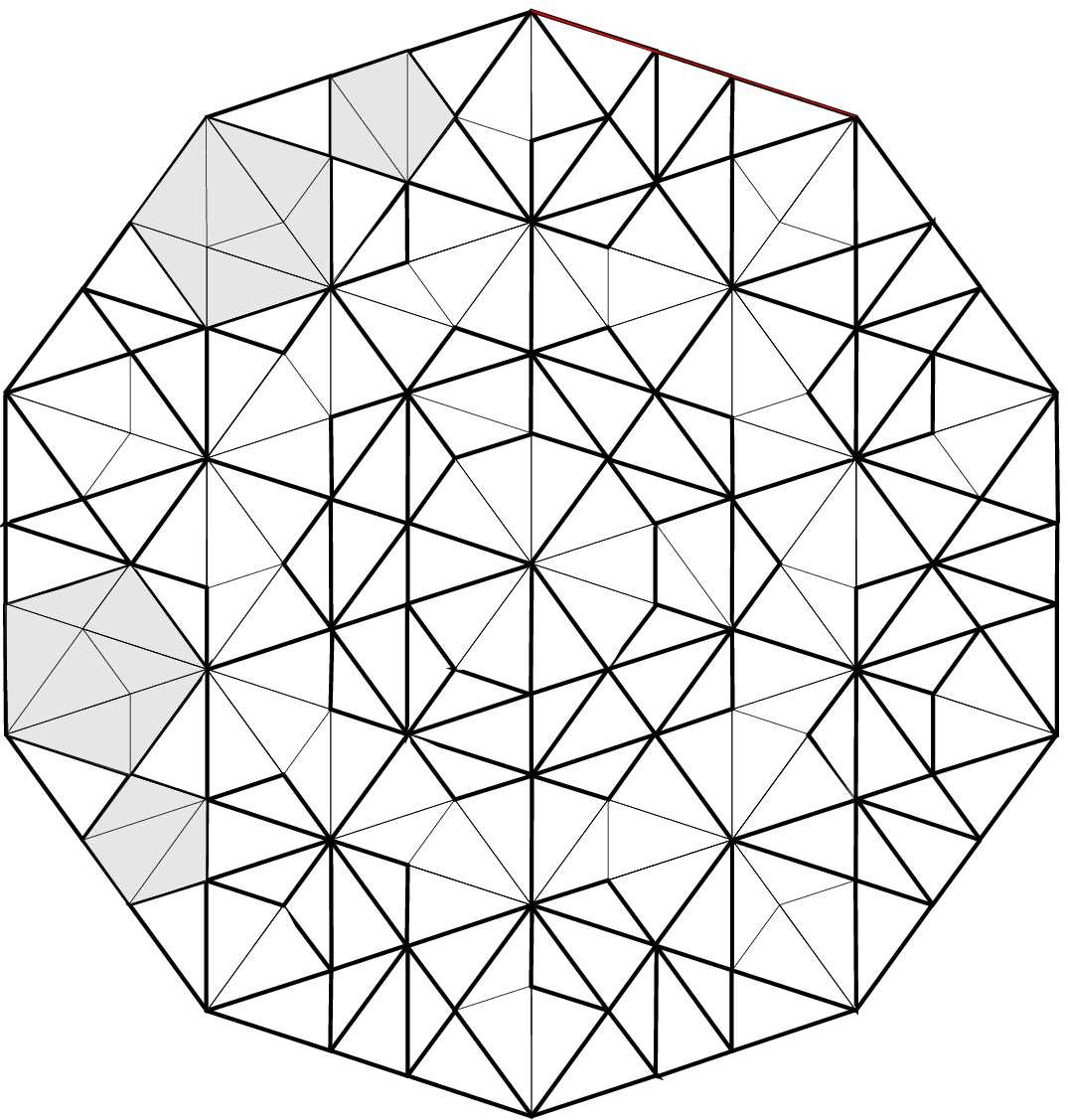}%
\end{picture}%
\setlength{\unitlength}{1243sp}%
\begingroup\makeatletter\ifx\SetFigFont\undefined%
\gdef\SetFigFont#1#2#3#4#5{%
  \reset@font\fontsize{#1}{#2pt}%
  \fontfamily{#3}\fontseries{#4}\fontshape{#5}%
  \selectfont}%
\fi\endgroup%
\begin{picture}(16264,17059)(1871,-18207)
\end{picture}

\end{center}

Fig. 5. Part of a triangle tiling $({\cal T}^*, A_4)$. The small and large pentagons 
(heavy edge lines) mark  the  linked Delone 
$D^a_{\parallel}, D^b_{\parallel}$-clusters which cover the tiling
by triangles (weak edge lines if not edge of a pentagon).
The four grey pentagons together form a fundamental domain of definition 
${\cal F}({\cal T}^*, A_4)$
for functions compatible with the triangle tiling. They comprise the two 
triangles each in $10$ orientations. 
 
\vspace{1cm}

{\bf 7 Prop.} The $D^a_{\parallel}, D^b_{\parallel}$-clusters of the triangle tiling are 
two types of
linked pentagons. Together with their mirror images 
they form a fundamental domain 
${\cal F}({\cal T}^*, A_4)$ for the triangle tiling. These Delone clusters
form a covering of the triangle tiling.

{\em Proof}: To determine configurations of triangles belonging to $D$-clusters
we select codings for vertex configurations 
by  Penrose tiles in the decagon 
which share a single (shallow or deep hole) 
vertex of the Voronoi cell. It turns out that, in the enumeration 
of \cite{BA},  the vertices 
$4,5,6,7$ produce small pentagonal $D^a_{\parallel}$-clusters and the 
vertex $2$
produces large $D^b_{\parallel}$-clusters. These two types of pentagons are 
linked and yield a covering of the triangle pattern as shown in Fig. 5.
The windows for the two pentagons can be found from the combination 
of vertex windows as shown in Fig. 6. Projections of shallow and deep
holes are marked as full and white circles. More details of the
Delone windows are given in appendix A. The proof of the covering property from
the projection and window method is given in appendix B. $\Box$
 
The triangle tiling is applied
to decagonal $AlCuCo$ in \cite{KR6}. The implications of linked pentagonal
$D$-clusters in terms of shared dual boundaries should be studied.

\begin{center}
\begin{picture}(0,0)%
\epsfig{file=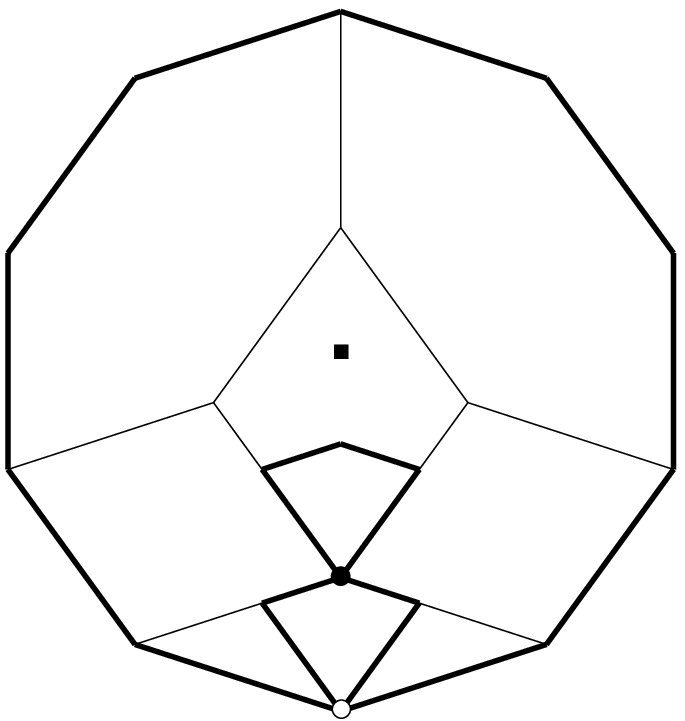}%
\end{picture}%
\setlength{\unitlength}{3315sp}%
\begingroup\makeatletter\ifx\SetFigFont\undefined%
\gdef\SetFigFont#1#2#3#4#5{%
  \reset@font\fontsize{#1}{#2pt}%
  \fontfamily{#3}\fontseries{#4}\fontshape{#5}%
  \selectfont}%
\fi\endgroup%
\begin{picture}(3870,4080)(688,-3558)
\put(2521,-2311){\makebox(0,0)[lb]{\smash{\SetFigFont{12}{14.4}{\familydefault}{\mddefault}{\updefault}$a$}}}
\put(2521,-3121){\makebox(0,0)[lb]{\smash{\SetFigFont{12}{14.4}{\familydefault}{\mddefault}{\updefault}$b$}}}
\end{picture}

\end{center}

Fig. 6. The windows $w(D^a), w(D^b)$ for the small and large
pentagonal $D$-clusters in the triangle tiling are two 
regions marked $(a, b)$ in the decagonal window $V_{\perp} \in E_{\perp}$
modulo 5fold rotations. For details of the construction compare 
appendix A.

\section{$V$- and $D$-clusters in icosahedral tilings.}

The three icosahedral modules $P, F, I$ can be projected from
the $P, F, I$ centered hypercubic lattice in $E^6$. All 
known icosahedral tilings are related to the canonical ones based
on duality.
From the analysis of the canonical icosahedral tilings \cite{KR1, KR5}
for the $P, F$ lattices one can immediately draw the following conclusion
on the shape of the $V$- and $D$-clusters in position space:

{\bf P lattice}: The Voronoi and Delone cells of the hypercubic 
lattice coincide in shape and so do their projections. 
The tiles are the well-known thick and thin rhombohedra of edge length
\ffo. Both the 
windows $V_{\perp}$ and 
the $D$-clusters $D_{\parallel}$ are the triacontahedra of Kepler
with edge length  \ffo.
From the translation orbits \cite{KR3} 
one can explore their relation to the fundamental domain 
${\cal F}({\cal T}^*, P)$.

{\bf F lattice}: The lattice $F$ has three different translation 
orbits of Delone cells denoted as $D^a, D^b, D^c$. 
In the tiling $({\cal T}^*, F)$, the window 
$V_{\perp}$ is again the triacontahedron of Kepler of edge length \ffo. 
The tiles are six tetrahedra projected from 3-boundaries of the Delone cells.
The vertex configurations were studied in \cite{KR5}. 
This tiling is used for modelling atomic positions of $AlPdMn$
and is closely related to the approach of Katz and Gratias \cite{KA1, KA2}.
From the present analysis we get three Delone clusters 
$D^a_{\parallel}, D^b_{\parallel}, D^c_{\parallel}$. 
Two of them are icosahedra with the edge lengths 
\zfo, $\tau$\zfo, where \zfo = $\frac{2}{\sqrt{\tau+2}}$ \ffo,
one is a dodecahedron of edge length \zfo, see Fig. 7. 
The windows for these Delone $D$-clusters and their relation
to the fundamental domain 
${\cal F}({\cal T}^*, F)$ must be analyzed.

In the dual tiling $({\cal T}, F)$, the three Delone windows are 
$D^a_{\perp}, D^b_{\perp}, D^c_{\perp}$.
The tiles are two rhombohedra and four pyramids projected from 
3-boundaries of the 
Voronoi domains. The Voronoi clusters
$V_{\parallel}$ are Kepler triacontahedra related to the fundamental 
domain ${\cal F} ({\cal T}, F)$.

\begin{center}
\begin{picture}(0,0)%
\epsfig{file=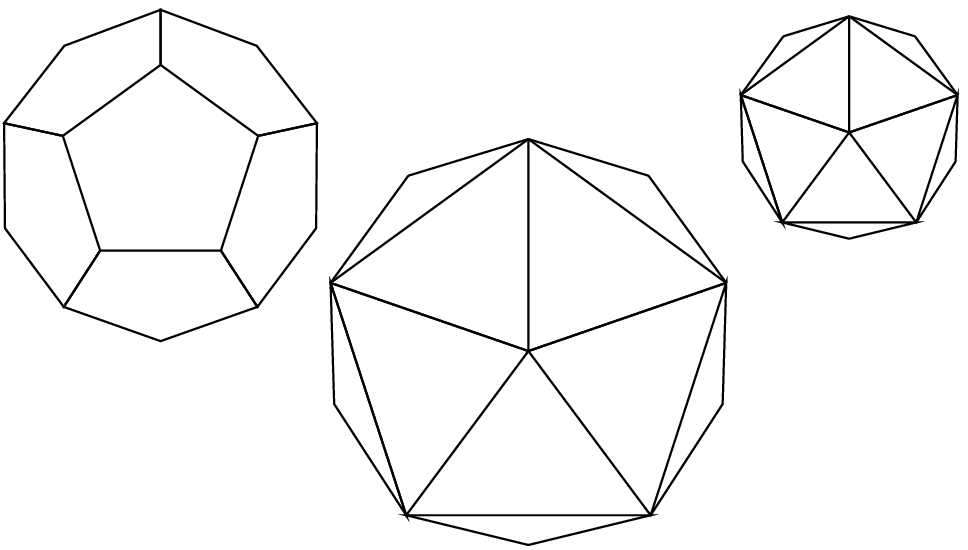}%
\end{picture}%
\setlength{\unitlength}{1243sp}%
\begingroup\makeatletter\ifx\SetFigFont\undefined%
\gdef\SetFigFont#1#2#3#4#5{%
  \reset@font\fontsize{#1}{#2pt}%
  \fontfamily{#3}\fontseries{#4}\fontshape{#5}%
  \selectfont}%
\fi\endgroup%
\begin{picture}(14596,8221)(665,-8271)
\put(691,-6181){\makebox(0,0)[lb]{\smash{\SetFigFont{20}{24.0}{\familydefault}{\mddefault}{\updefault}$c$}}}
\put(4111,-7861){\makebox(0,0)[lb]{\smash{\SetFigFont{20}{24.0}{\familydefault}{\mddefault}{\updefault}$a$}}}
\put(10441,-811){\makebox(0,0)[lb]{\smash{\SetFigFont{20}{24.0}{\familydefault}{\mddefault}{\updefault}$b$}}}
\end{picture}

\end{center}

Fig. 7. Schematic view of the three Delone $D$-clusters  for
the icosahedral  tiling $({\cal T}^*, F)$:
The three Delone
$D^a, D^b, D^c$-clusters are a dodecahedron and two icosahedra.
\vspace{1cm}

Structure questions to be studied are the fundamental domain and
covering property,  
the linkage of $V, D$-clusters in relation to the modules and shared 
dual boundaries, 
their local point symmetry, and atomic positions
on them. In all cases the technique of windows for  the tilings, 
the tiles and vertex 
configurations, and for atomic positions  is available. 
It allows to implement the project 
of $D$- and $V$-clusters and coverings in the structure theory
and in the physics  of icosahedral quasicrystals.  

\section{Summary and outlook.}
The covering project analyzed here for dual tilings from lattice 
embedding has the following features:\\
(1) {\em Fundamental domains}: The analysis depends crucially on the
notion Def. 2 of a fundamental domain for functions on the position space 
$E_{\parallel}$ compatible with the tiling.\\
(2) {\em Clusters}: The clusters should 
be related to the  fundamental
domain for functions compatible with the tiling. This relation is given
for Voronoi or Delone clusters in the 2D Penrose and triangle tilings 
and must be explored for the dual 3D icosahedral tilings.\\
(3) {\em Center positions and their windows}: 
The center positions are assumed to be for Voronoi clusters
the parallel projected lattice point positions, for Delone clusters the 
parallel projected hole positions
(vertices of Voronoi cells) whose perpendicular projections fall into
the window(s) of the tiling. This property is verified for the 
1D Fibonacci and $A_2$ -based tiling and for the dual 2D Penrose and triangle 
tilings. In all these cases the positions and windows  
for the centers of the clusters are explicitly determined.\\
(4) {\em Cluster recognition and linkage in the tilings}: 
In the Fibonacci, the $A_2$-based and triangle tiling the clusters
appear as collections of full tiles. Their possibly alternative
compositions from these tiles  and their linkage and  overlaps 
can be unified by appropriate subdivision 
with or (in case of the $A_2$-based tiling) without the use of deflation.
For the Penrose rhombus tiling one needs one level of deflation
plus the Robinson subdivision to recognize the decagon clusters and 
to analyze their overlaps and linkage.\\
(5) {\em Covering property}: For Voronoi and Delone clusters
in 1D tilings with the assumption (3) on  their centers, the 
covering property can be verified 
in an elementary way. For  the dual 
2D tilings the covering property  can be shown from the explicit projection 
and window technique (example in appendix B). \\
(6) {\em Icosahedral clusters}: For the dual 3D icosahedral tilings the
Voronoi and Delone clusters are known. The window technique is 
more involved but available. 
It must be implemented to relate them to fundamental domains (2), 
to find  the cluster centers and their windows 
according to (3), to study   the covering property (5), and to find the
linkage of clusters.

\newpage
\section*{Appendix A: Windows for the pentagonal covering.}

We present an independent analysis of the windows for the 
centers of pentagonal Delone clusters in the tiling $({\cal T}^{*}, A_4)$
of section 5, using the geometry of \cite{BA}. The 
acute and obtuse triangle tiles $t_1, t_2$ of this tiling 
in $E_{\parallel}$ have as windows $w(t_1), w(t_2)$ in $E_{\perp}$
the dual thick and thin Penrose rhombus tiles appearing
within the decagonal Voronoi window $V_{\perp}$. 
Vertex configurations in the tiling are coded in this
window by intersections of the rhombus windows. For the
complete list of vertices and windows we refer to \cite{BA}. In
Fig. 8 we show the collection of those tiles and their windows which
together determine the windows for the Delone clusters
$D^a_{\parallel}, D^b_{\parallel}$.

\begin{center}
\begin{picture}(0,0)%
\epsfig{file=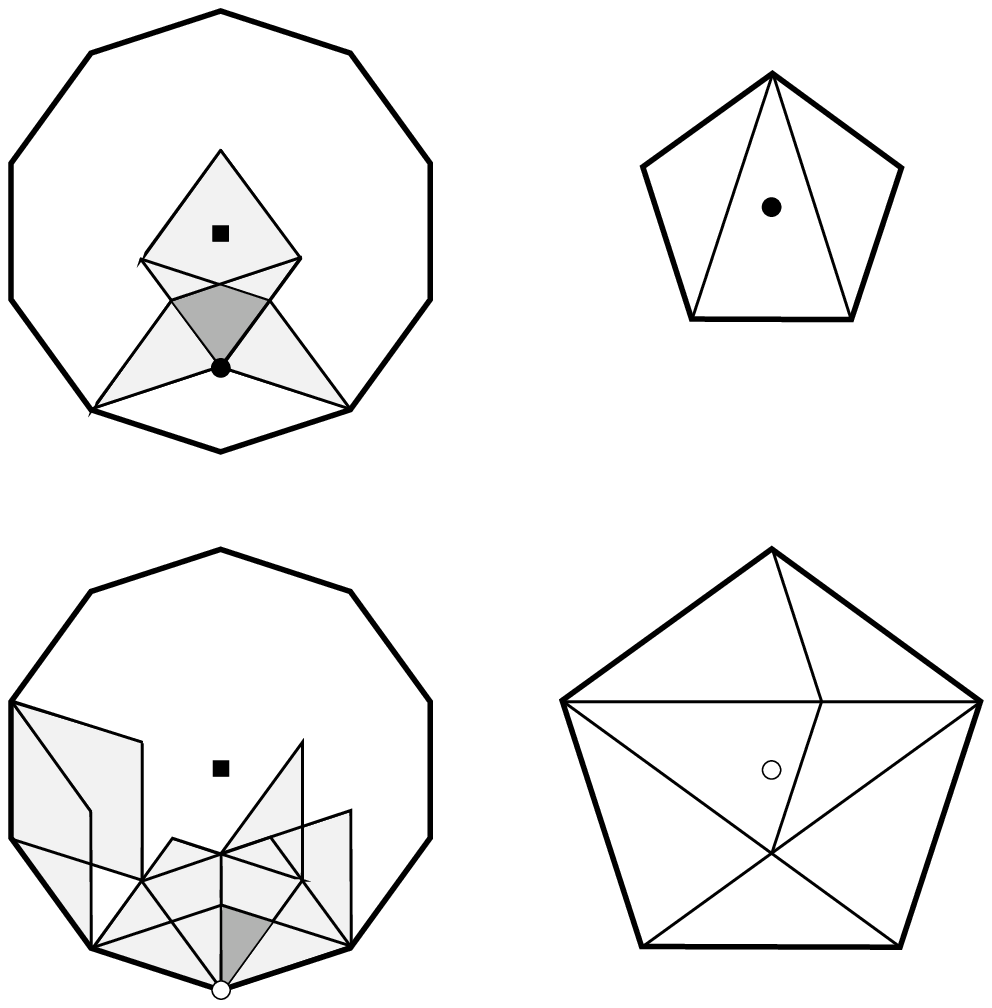}%
\end{picture}%
\setlength{\unitlength}{3315sp}%
\begingroup\makeatletter\ifx\SetFigFont\undefined%
\gdef\SetFigFont#1#2#3#4#5{%
  \reset@font\fontsize{#1}{#2pt}%
  \fontfamily{#3}\fontseries{#4}\fontshape{#5}%
  \selectfont}%
\fi\endgroup%
\begin{picture}(5951,6801)(676,-6382)
\put(676,-6324){\makebox(0,0)[lb]{\smash{\SetFigFont{12}{14.4}{\familydefault}{\mddefault}{\updefault}
$E_{\perp}$
}}}
\put(3826,-6316){\makebox(0,0)[lb]{\smash{\SetFigFont{12}{14.4}{\familydefault}{\mddefault}{\updefault}
$E_{\parallel}$
}}}
\put(676,-2898){\makebox(0,0)[lb]{\smash{\SetFigFont{12}{14.4}{\familydefault}{\mddefault}{\updefault}
$w(D^b)$
}}}
\put(3856,-2898){\makebox(0,0)[lb]{\smash{\SetFigFont{12}{14.4}{\familydefault}{\mddefault}{\updefault}
$D^b_{\parallel}$
}}}
\put(676,224){\makebox(0,0)[lb]{\smash{\SetFigFont{12}{14.4}{\familydefault}{\mddefault}{\updefault}
$w(D^a)$
}}}
\put(3826,224){\makebox(0,0)[lb]{\smash{\SetFigFont{12}{14.4}{\familydefault}{\mddefault}{\updefault}
$D^a_{\parallel}$
}}}
\end{picture}

\end{center}

Fig. 8 The windows  for the Delone $D$-clusters of the pentagonal 
covering of section section 5 modulo the 
point group $D_{10}$. The small pentagon 
$D^a_{\parallel} \in E_{\parallel}$ (top, right)
is formed from $1$ acute triangle $t_1$ and $2$ obtuse triangles $t_2$. 
The windows 
for these tiles are $1$ thick and $2$ thin rhombus tiles 
$w(t_1), w(t_2)$ in the 
decagonal window $V_{\perp} \in E_{\perp}$ (top, left). Their
intersection is the window $w(D^a)$ for a vertex position of 
the small Delone cluster (grey).
The large pentagon $D^b_{\parallel} \in E_{\parallel}$ (bottom, right)
is formed by $4$ acute triangles $t_1$  and $3$ obtuse triangles $t_2$. 
The windows are $4$ thick and $3$ thin rhombus tiles in the decagonal 
window (bottom, left). $3$ thick and $2$ thin rhombus tiles (light grey)
intersect in half of the window $w(D^b)$ (grey) which codes the lower
vertex position of the large Delone cluster. The remaining acute and obtuse 
triangle at the upper vertex of the pentagon (right) are forced by 
the remaining thin and thick rhombus  windows 
shown in the decagon (light grey, left). A reflection in a vertical line
both in $E_{\parallel}$ and $E_{\perp}$ yields a second version of the
cluster $D^b_{\parallel}$ and the second half  of the window 
$w(D^b)$ shown in Fig. 6.

\section*{Appendix B: Covering from  projection and windows.}

Given a tiling ${\cal T}$ and a set of clusters with their center positions,
the {\em covering property}  requires that all parts of any tile from the tiling
be covered by at least one of the clusters.

For the decagon covering of the
Penrose tiling $({\cal T}, A_4)$ the covering property was shown by Gummelt \cite{GU}.
It was shown in section 4 that the $V$-clusters for the Penrose
tiling coincide with the decagons of Gummelt. The covering 
property holds by this coincidence. In what follows we give a constructive proof from projections and windows 
for the covering property of the triangle tiling by pentagonal
Delone clusters. A similar constructive proof applies  for the Penrose tiling.
W 

{\bf 8 Prop.} The pentagonal Delone clusters 
$D^a_{\parallel}, D^b_{\parallel}$ with
the center positions described in section 5 form  a covering of
the triangle tiling $({\cal T}^{*}, A_4)$. 

{\em Proof}: The tiles of this tiling are acute or obtuse golden 
triangles $t_1, t_2$,
compare Figs. 5, 8 and 9.
Their dual windows have the  thick and  thin Penrose rhombus shape
$w(t_1), w(t_2)$ and are located 
in the decagonal Voronoi window $V_{\perp}$. On each rhombus we mark
its shallow hole vertex by a full circle. For a fixed orientation
of $t_1, t_2$, each window $w(t_1), w(t_2)$ appears in the decagon
in three shifted positions. These three positions are shown
in Fig. 9. 
  
\begin{center}
\begin{picture}(0,0)%
\epsfig{file=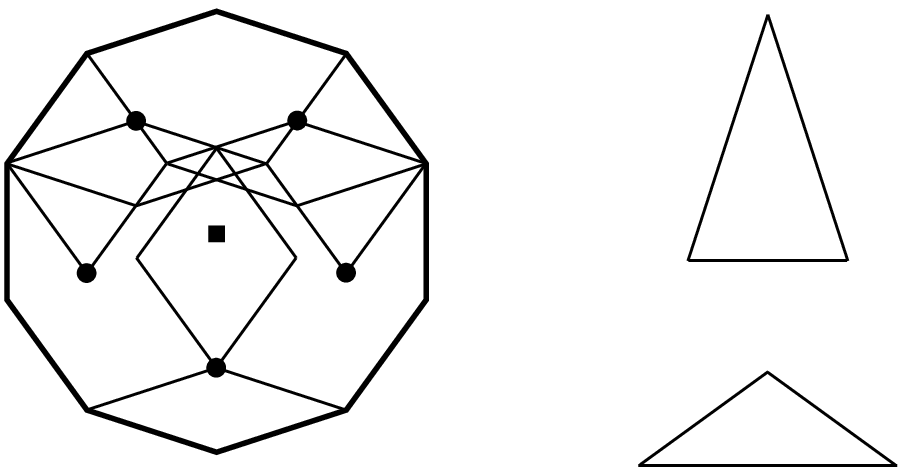}%
\end{picture}%
\setlength{\unitlength}{3315sp}%
\begingroup\makeatletter\ifx\SetFigFont\undefined%
\gdef\SetFigFont#1#2#3#4#5{%
  \reset@font\fontsize{#1}{#2pt}%
  \fontfamily{#3}\fontseries{#4}\fontshape{#5}%
  \selectfont}%
\fi\endgroup%
\begin{picture}(5142,3052)(1020,-2705)
\put(1070,-2647){\makebox(0,0)[lb]{\smash{\SetFigFont{12}{14.4}{\familydefault}{\mddefault}{\updefault}$E_{\perp}$}}}
\put(4648,-2647){\makebox(0,0)[lb]{\smash{\SetFigFont{12}{14.4}{\familydefault}{\mddefault}{\updefault}$E_{\parallel}$}}}
\put(4505,-1827){\makebox(0,0)[lb]{\smash{\SetFigFont{12}{14.4}{\familydefault}{\mddefault}{\updefault}$t_2$}}}
\put(4505, 78){\makebox(0,0)[lb]{\smash{\SetFigFont{12}{14.4}{\familydefault}{\mddefault}{\updefault}$t_1$}}}
\end{picture}

\end{center}

Fig. 9. 
The acute and obtuse triangle 
$t_1, t_2$ (right) of
the triangle tiling $({\cal T}, A_4) \in E_{\parallel}$.  
Their  windows $w(t_1), w(t_2)$ in $E_{\perp}$ (left) for fixed orientations
are dual thick and thin Penrose rhombus tiles in the decagon 
$V_{\perp}$. The shallow hole vertex is marked on each rhombus window
by a full circle.
Each rhombus window $w(t_i)$ appears in three shifted  positions. 
\vspace{1cm}

For all other orientations under the action of the point group $D_{10}$
in $E_{\parallel}$,
the windows transform under the corresponding action of
$D_{10}$ but now in $E_{\perp}$ \cite{BA}. For example under a rotation of the tiles
by $2\pi/5$ their windows must be rotated by $4\pi/5$.
 
Consider next the Delone clusters $D^a_{\parallel}, D^b_{\parallel}$
of section 5 with their composition out of   $3$ and $7$ triangle tiles
respectively. 
In appendix A Fig.7 we give for both clusters with fixed
orientation the collection 
of the $3$ and $7$  rhombus windows $w(t_1), w(t_2)$ 
within $V_{\perp}$. A comparison of 
the left-hand sides of Figs. 7 and 8 for $E_{\perp}$ shows: Any single 
rhombus window $w(t_i)$ for a single triangle tile $t_i$ 
in  the $3$ possible shifted positions of Fig. 8
can be brought modulo $D_{10}$ into coincidence with at least
one rhombus window $w(t_j)$ from the collection for $D^a_{\parallel}$ or
$D^b_{\parallel}$ given in Fig. 7. In $E_{\parallel}$ this implies that
for any triangle tile $t_i \in ({\cal T}, A_4)$ one can construct at 
least  one Delone cluster 
$D^a_{\parallel}$ or $D^b_{\parallel}$ that covers this tile. $\Box$

\end{document}